\documentclass[%
 reprint,
 amsmath,amssymb,
 aps,
prd,
]{revtex4-2}

\usepackage{graphicx}
\usepackage{dcolumn}
\usepackage{bm}
\usepackage{xcolor} 
\usepackage{hyperref}
\usepackage{enumitem}
\def\be{\begin{equation}}
\def\ee{\end{equation}}
\def\bea{\begin{eqnarray}}
\def\eea{\end{eqnarray}}

\begin{document}

\preprint{APS/123-QED}

\title{Bias in the tensor-to-scalar ratio from self-interacting dark radiation}

\author{Nahuel Mirón-Granese}
\email{nahuelmg@df.uba.ar}

\affiliation{Consejo Nacional de Investigaciones Científicas y Técnicas (CONICET),
Godoy Cruz 2290, Ciudad de Buenos Aires C1425FQB, Argentina}
\affiliation{Universidad Nacional de La Plata, Facultad de Ciencias Astronómicas y Geofísicas, Paseo del Bosque, La Plata B1900FWA, Buenos Aires, Argentina }
\affiliation{Universidad de Buenos Aires, Facultad de Ciencias Exactas y Naturales, Departamento de Física,\\ Intendente Güiraldes 2160, Ciudad de Buenos Aires C1428EGA, Argentina}

\author{Claudia G. Scóccola}%
 \email{claudia.scoccola@uchile.cl}
\affiliation{%
 Departamento de F\'isica, FCFM, Universidad de Chile, Blanco Encalada 2008, Santiago, Chile}%




\date{\today}

\begin{abstract}

We investigate the cosmological imprint of self-interacting dark radiation (DR) on the primordial $B$-mode angular power spectrum and its impact on the estimation of the tensor-to-scalar ratio $r$. We consider a minimal model in which DR is described as an effectively massless axion-like particle with quartic self-interactions. These interactions are incorporated into the Einstein–Boltzmann equations using the relaxation time approximation and implemented in the \texttt{CLASS} code. We show that increasing the strength of DR self-interactions suppresses anisotropic stress, thereby reducing the damping of gravitational waves and leading to an enhancement of the primordial $B$-mode signal relative to the free-streaming case. Using mock CMB data and Markov Chain Monte Carlo analyses, we show that neglecting DR self-interactions may bias the inferred value of $r$ by an amount comparable to the uncertainty expected in forthcoming CMB polarization experiments, such as the ground-based \textit{Simons Observatory} and the satellite missions \textit{LiteBIRD} and PICO. Our results emphasize the importance of properly modeling DR interactions in future precision searches for primordial $B$-modes in order to obtain unbiased constraints on inflationary gravitational waves.
\end{abstract}

\maketitle


\section{\label{sec:intro}Introduction}

Primordial $B$-mode polarization of the cosmic microwave background (CMB) provides a unique observational window onto inflationary gravitational waves (GW) and hence the energy scale of inflation \cite{weinberg_cosmology,thefirstthreeseconds,tensorsandbmodes_review,inflationtheoryobservations}. A robust detection (or constraint) of the tensor-to-scalar ratio $r$ would discriminate among large classes of inflationary models and sharpen our understanding of the very early Universe. Achieving this goal requires high control of instrumental systematics, astrophysical foregrounds, and the microphysics of the primordial plasma that can modify the propagation and imprint of tensor perturbations on the CMB.

It is well known that the coupling between ultra-relativistic (UR) species present in the primordial plasma and the GW modifies the primordial $B$-mode spectrum of the CMB polarization \cite{Polnarev_1985,weinberg04,watanabe06}, thereby affecting the observable tensor-to-scalar ratio today. UR species include photons and those non-photonic relativistic components commonly grouped as neutrinos and possible beyond-Standard-Model relics \cite{lr_dvorkin2022physicslightrelics,lr_Ge_2023,lr_Giovanetti_2022}, referred to as dark radiation (DR). After decoupling from the plasma, UR species are typically treated as free-streaming: in that limit they develop a nonvanishing anisotropic stress which back-reacts on the GW evolution producing a damping of the amplitude of the GW spectrum. Conversely, when interactions among UR species are efficient they isotropize the distribution function, suppress anisotropic stress and thus remove the GW damping. The competition between the species’ interaction rate and the Hubble expansion rate therefore determines whether the species behave as a free-streaming gas or as a fluid-like component, with direct consequences for the amplitude and shape of the primordial GW spectrum \cite{nahuelesteban,nahuel,baym,loverde2022}. The current GW spectrum including these effects was explicitly computed in \cite{nahuel, loverde2022}, while the effect on the primordial $B$-mode spectrum is studied in this paper.

The key question motivating this work is whether self-interactions among UR species, and in particular among a cosmologically relevant DR component, can alter the observed $B$-mode signal sufficiently to bias the inference of $r$. While current CMB and large-scale-structure measurements place strong limits on many classes of neutrino interactions at the epochs probed by the primary CMB anisotropies \cite{neutrinosfreestreaming,Choudhury_2021,Choudhury_2022,Camarena_2025,neu_Allali_2024,neu_poudou2025selfinteractingneutrinoslightrecent,neu_He_2024,neu_Aloni_2023,neu_Camarena_2023,neu_Forastieri_2019,neu_Park_2019,neu_Lancaster_2017}, models of new self-interacting DR remain phenomenologically viable. In these scenarios the DR abundance, parametrized by $\Delta N_{\rm eff}$, is tightly constrained by data \cite{planck18,neff-spt25,neff-act25,Allali_2025,Saravanan_2025}, whereas the possible types of interactions and their strengths can still cover a broad parameter space without being excluded \cite{drint,Trendafilova_2025,Bagherian_2025,Buen-Abad2025,idr_Sch_neberg_2023,idr_Ghosh_2022,idr_Blinov_2020,idr_Buen_Abad_2018,dm-dr_Archidiacono_2019,dm-dr_Choi_2021}.

As a concrete and well-motivated benchmark we adopt a self-interacting axion-like particle (ALP), described as an effectively massless scalar field with a quartic self-interaction potential $\lambda\phi^4/4!$. For cosmological purposes this model is fully specified by the DR abundance $\Delta N_{\rm eff}$ and the dimensionless self-coupling constant $\lambda$. Reference \cite{drint} performed a dedicated parameter study of this model against Planck 2018 primary temperature and polarization data, CMB lensing, and BAO, finding that cosmological datasets tightly restrict $\Delta N_{\rm eff}$ while allowing substantial freedom in $\lambda$. Based on these results, we evaluate different representative scenarios within the observationally allowed region, focusing on $\Delta N_{\rm eff}\lesssim 0.5$ and self-couplings in the range $\lambda\sim 10^{-12}\!-\!10^{-9}$, as detailed in Sec. \ref{sec:impactoncls}.

In this paper we explore the observable signatures of such self-interacting DR on primordial $B$ modes and we quantify whether neglecting DR self-interactions in the analysis model can bias estimates of $r$. The physical mechanism is: self-interactions suppress the anisotropic stress of the DR, thereby reducing the GW damping and leading to an enhancement of power in the primordial $B$-mode spectrum relative to the free-streaming case. If data are analyzed assuming free-streaming DR when in fact DR self-interacts, the recovered tensor amplitude can be systematically shifted with a magnitude that depends on $\Delta N_{\rm eff}$ and the self-interaction strength $\lambda$. We therefore ask if the induced bias on $r$ might be comparable to the statistical uncertainties expected for near-future experiments.

To address this issue, we incorporate the DR self-interaction terms into the full Einstein--Boltzmann system and compute the tensor-sector transfer functions with a modified version of \texttt{CLASS} (see Sec. \ref{sec:modelimplementation}). We then generate realistic mock primordial (unlensed) $B$-mode spectra including instrumental noise and cosmic variance, and perform Markov Chain Monte Carlo (MCMC) parameter estimation under fitting models that either include or neglect DR interactions. This pipeline allows us to (i) map the scale- and parameter-dependent modifications of the $B$-mode spectrum induced by interacting DR (Sec. \ref{sec:impactoncls}) and (ii) quantify the potential systematic bias on the recovered tensor-to-scalar ratio $\Delta r\equiv(r_{\rm best-fit}-r_{\rm fid})/r_{\rm fid}$ for representative experimental configurations (Sec. \ref{sec:impactonr}).

The current experimental landscape of CMB polarization includes imaging telescopes such as BICEP/Keck \cite{Hui2018,bicepkeck}, CLASS \cite{2022ApJ...926...33D}, ACT \cite{2025arXiv250314452L}, and SPT-3G \cite{2021PhRvD.104b2003D}, which are already delivering high-quality with the most stringent constraint on the tensor-to-scalar ratio to date $r < 0.032$ at 95\% CL by BICEP/Keck and Planck PR4 \cite{Tristram_2022,bicepkeck}. QUBIC \cite{QUBICI,QUBICII}, as the first bolometric interferometer telescope, provides a complementary approach to these measurements. Upcoming experiments, such as \textit{The Simons Observatory} (SO) and the \textit{LiteBIRD} satellite, are designed to reach sensitivities to primordial $B$-modes at the level of $r \sim 10^{-3}$.
SO will deploy three Small Aperture Telescopes (SATs) and one Large Aperture Telescope (LAT) in the Atacama Desert, with a total of nearly 60,000 TES bolometers spanning six frequencies from 27 to 280\,GHz \cite{Ade2019,so_2025}. The SATs will survey approximately 10\% of the sky with a projected white noise level of $2\,\mu{\rm K}$-arcmin in combined 93/145\,GHz bands, enabling a constraint on the tensor-to-scalar ratio with an expected sensitivity of $\sigma(r)\simeq 0.003$. The LAT will cover about 40\% of the sky with arcminute resolution and $6\,\mu{\rm K}$-arcmin noise in the same bands. Complementing these ground-based measurements, \textit{LiteBIRD} (scheduled for launch to L2 around 2032) will perform a full-sky survey over three years with 15 frequency bands between 34 and 448\,GHz. Its typical angular resolution is $\sim 0.5^\circ$ at 100\,GHz, with a target sensitivity of $2.2\,\mu{\rm K}$-arcmin in polarization, designed to achieve $\delta r \simeq 0.001$ including systematic uncertainties \cite{Allys2022,giardiello_requirements_2025}. Together, the deep and high-resolution observations of SO and the all-sky coverage and systematic control of \textit{LiteBIRD} will provide a decisive test of inflationary models.
Beyond SO and \textit{LiteBIRD}, the Probe of Inflation and Cosmic Origins (PICO) is a NASA probe-class mission concept designed to map the full sky with an unprecedented sensitivity of $0.61\,\mu{\rm K}$-arcmin across 21–799\,GHz, enabling constraints on inflationary physics, neutrino masses, structure formation, and Galactic foregrounds \cite{2018SPIE10698E..46Y,Hanany2019,Aurlien_2023}.

Given these experimental goals, even modest systematic shifts in $r$ arising from unmodeled physics could affect the interpretation of any claimed detection. We therefore map the parameter space $(\Delta N_{\rm eff},\lambda)$ and identify the regions where DR self-interactions should be included in analysis pipelines to avoid biased inferences of primordial GW.

The remainder of the paper is organized as follows. In Section \ref{sec:modelimplementation} we introduce the self-interacting DR model and its implementation in the Einstein--Boltzmann solver. Section \ref{sec:impactoncls} presents the impact of DR self-interactions on the primordial $B$-mode spectrum, while Section \ref{sec:impactonr} quantifies the induced bias on the tensor-to-scalar ratio for a realistic mock spectra of future CMB polarization experiments. Finally, Section \ref{sec:conclusions} summarizes our findings and discusses their implications for future measurements of primordial gravitational waves.

\section{DR model implementation}\label{sec:modelimplementation}

We include the dynamics of the DR in the Einstein-Boltzmann equations for both the background and the linear perturbations through a modified version of the well-known Einstein-Boltzmann solver called \texttt{CLASS} \cite{class_1,class_2}.

For the background evolution, DR can be treated as an additional radiation-like component whose abundance is fixed by $\Delta N_{\rm eff}$. Its energy density relative to photons is given by
\begin{equation}
\frac{\rho_{\rm DR}}{\rho_\gamma} 
= \frac{g_{\rm DR}}{2} \left( \frac{T_{\rm DR}}{T_\gamma} \right)^4
= \frac{7}{8} \left( \frac{4}{11} \right)^{4/3} \Delta N_{\rm eff},
\label{energydensity}
\end{equation}
and the corresponding present-day temperature reads
\begin{equation}
T_{\rm DR,0} 
= \left[ \frac{7}{8} \left( \frac{4}{11} \right)^{4/3} \Delta N_{\rm eff} \right]^{1/4} T_{\gamma,0},
\label{temperature}
\end{equation}
where $T_{\gamma,0}$ denotes the current photon temperature and $g_{\rm DR}=2$ is the number of internal degrees of freedom.

This component was thermally produced in the very early Universe and rapidly decoupled from the primordial plasma at some high temperature $T_{\rm DR,\,dec}$. That decoupling epoch sets its abundance $\Delta N_{\rm eff}$ and current temperature $T_{\rm DR,0}$ through Eqs. (\ref{energydensity})--(\ref{temperature}), in close analogy with neutrinos. After decoupling, self-interactions, when efficient, do not re-establish thermalization with photons but instead they turn the DR distribution function into its equilibirum form erasing the fluctuations, as discussed in Sec. \ref{sec:impactoncls}.

The particular model of DR considered in this work is detailed in Appendix B of \cite{drint} and it is effectively described by a massless ALP, $\phi$, with a self-interacting Lagrangian
\bea
\mathcal{L}_{\rm int}=-\lambda\, \phi^4/4!\,.\label{lint}
\eea
The general and usual framework for including such interactions in the Boltzmann equations is through the collision integral using the relaxation time approximation \cite{Oldengott_2015,Oldengott_2017}. Within this approximation, the collision integral is proportional to the product of the thermally averaged interaction rate, $\langle\Gamma\rangle$, and a numerical coefficient $\alpha$, which we refer to as the relaxation time coefficient. This coefficient depends on the specific interaction type and must be computed accordingly. In \cite{drint}, particularly in Appendix B, the authors present the computation of both the thermally averaged interaction rate and the relaxation time coefficient for the DR model considered in this work.

The thermal-averaged self-interaction rate $\langle\Gamma\rangle$ can be roughly estimated on dimensional grounds. Indeed $\langle\Gamma\rangle\sim n \langle\sigma v\rangle$ where $n$ is the number density and the average cross section $\langle\sigma v\rangle\sim |\mathcal{M}|^2/E^2$ with $|\mathcal{M}|$ and $E$ the scattering amplitude and the energy scale of the considered scattering, respectively. For the $2\leftrightarrow 2$ scattering with the interaction Lagrangian of Eq. (\ref{lint}), at the tree level order within the ultra-relativistic regime, we get $n\sim T^3$, $|\mathcal{M}|\sim\lambda$, $E\sim T$, and thus $\langle\Gamma\rangle\sim \lambda^2 T$, where $T$ is the physical temperature. Finally, this can be rewritten in terms of the scaling of the temperature with the scale factor of the universe $a$ and the current temperature $T_0$ as $\langle\Gamma\rangle\sim\lambda^2T_0/a$. An accurate computation \cite{drint} gives
\bea
\langle\Gamma\rangle=\frac{\pi\lambda^2T_{\rm DR,0}}{23040\,a}\,.
\label{gamma_lambda}
\eea
It is possible to define a constant comoving relaxation or collision time as
\bea
\tau = \frac{1}{a\,\langle\Gamma\rangle}=\frac{23040}{\pi\lambda^2T_{\rm DR,0}}\,,
\label{taugammalambda}
\eea
which is the characteristic time scale of the interaction.

\subsection{First-order perturbations}\label{sec:firstorderperturbations}

The first-order Einstein--Boltzmann system couples the metric to the energy--momentum tensor through Einstein equations, and the one-particle distribution function of the different species to the collision integral through the Boltzmann equations. Since our main interest is the impact on primordial CMB $B$-modes, it is sufficient to focus only on the evolution of the tensor sector within the scalar--vector--tensor (SVT) decomposition of cosmological perturbations, assuming a homogeneous and isotropic Friedmann--Lemaître--Robertson--Walker (FLRW) background \cite{tensorsandbmodes_zaldarriaga,tensorsandbmodes_kamionkowski,tensorsandbmodes_review,weinberg_cosmology}.

The energy--momentum tensor can be split into a background and a perturbed part,
\begin{equation}
T_{\mu\nu} = \bar{T}_{\mu\nu} + \delta T_{\mu\nu},
\end{equation}
where $\bar{T}_{\mu\nu}$ corresponds to the perfect-fluid background, while $\delta T_{\mu\nu}$ encodes fluctuations. It turns out that since the background is homogeneous and isotropic, the scalar, vector, and tensor sectors evolve independently at linear order. The components $\delta T_{00}$ and $\delta T_{0i}$ involve only scalar and vector modes, whereas $\delta T_{ij}$ contains also tensor contributions. Thus, to study tensor modes, it is sufficient to expand the spatial sector in terms of normal modes \cite{hu} as
\begin{equation}
\delta T^i{}_j = p \sum_m \pi^{(m)} Q^{(m)i}{}_j,
\end{equation}
where $p$ is the homogeneous pressure, $\pi^{(m)}$ is a scalar describing the anisotropic stress, and $Q^{(m)i}{}_j$ is the corresponding rank--2 normal mode. The index $m=0, \pm1, \pm2$ labels scalar, vector, and tensor contributions, respectively. Since our goal is to analyze primordial $B$-modes, we restrict to the tensor sector ($m=\pm2$) of the perturbed first-order Einstein--Boltzmann equations.

For tensor modes, the relevant fluctuations of the energy--momentum tensor are given by the transverse and traceless (TT) spatial projection of $\delta T^i{}_j$:
\begin{equation}
\big[\delta T^i{}_j\big]^{\rm TT} 
= p \Big[ \pi^{(2)} Q^{(2)i}{}_j + \pi^{(-2)} Q^{(-2)i}{}_j \Big].
\end{equation}

On the metric side, we expand around the conformally flat FLRW background as
\begin{equation}
g_{\mu\nu} = a^2 \big( \eta_{\mu\nu} + h_{\mu\nu} \big),
\end{equation}
where $a$ is the scale factor, $\eta_{\mu\nu}$ is the Minkowski metric, and $h_{\mu\nu}$ represents first-order perturbations. Restricting again to tensor modes, the TT part reads
\begin{equation}
h_{ij}^{\rm TT} = 2h^{(2)} Q^{(2)}_{ij} + 2h^{(-2)} Q^{(-2)}_{ij}.
\end{equation}

The evolution and initial conditions for the $m=2$ and $m=-2$ sectors are identical. For simplicity, we write down the equations only for $m=2$, and adopt the shorthand notation $h^{(2)} \equiv h$ and $\pi^{(2)}$.

We begin with the dynamics of tensor metric perturbations, i.e. gravitational waves $h$, which correspond to the TT projection of the spatial Einstein equations. In the notation of \cite{tram2013}, they read
\begin{equation}
h'' + 2\frac{a'}{a} h' + k^2 h 
= \sum_i 8\pi G\, a^2 p_i \, \pi^{(2)}_i,
\label{einsteinH}
\end{equation}
where $k$ is the comoving wavenumber, primes denote derivatives with respect to conformal time, and $p_i$ and $\pi^{(2)}_i$ are the background pressure and TT anisotropic stress, respectively, of species $i$.

For the Boltzmann equations, we include the dynamics of the one-particle distribution function of DR can be written as
\begin{equation}
f(\eta,\vec{x},\vec{n}) = f_0(\eta)\big[1+\Theta(\eta,\vec{x},\hat{n})\big],
\end{equation}
where $f_0(\eta)$ is the equilibrium distribution and $\Theta$ represents perturbations.  
As usual, we expand the perturbations in a basis of normal modes $G^m_\ell$ \cite{hu}:
\begin{equation}
\Theta(\eta,\vec{x},\vec{n}) = 
\int \frac{d^3k}{(2\pi)^3} 
\sum_{\ell}\sum_{m=-2}^{2} \Theta^{(m)}_\ell G^m_\ell.
\end{equation}
Restricting to the tensor sector ($m=2$), the Boltzmann hierarchy becomes
\begin{align}
\Theta'^{(2)}_2 &= -\frac{\sqrt{5}}{7}\,k\,\Theta^{(2)}_3 
- \alpha_2 a \langle \Gamma \rangle \Theta^{(2)}_2 
- \dot{h}, 
\label{boltzmannhierarchy} \\
\Theta'^{(2)}_\ell &= k \left[
\frac{s_\ell}{2\ell+1}\Theta^{(2)}_{\ell-1}
- \frac{s_{\ell+1}}{2\ell+3}\Theta^{(2)}_{\ell+1}
\right]
- \alpha_\ell a \langle \Gamma \rangle \Theta^{(2)}_\ell, \nonumber
\end{align}
with $s_\ell = \sqrt{(\ell^2-4)/\ell^2}$. The hierarchy is closed using the standard truncation described in \cite{ma95}, as implemented in \texttt{CLASS}.  
The relaxation-time coefficients $\alpha_\ell$ were computed in \cite{drint}; for simplicity, we adopt $\alpha_\ell = \alpha_2 \simeq 0.188$.

The TT projection of the DR anisotropic stress is related to the quadrupole of the perturbed distribution function as \cite{hu, tram2013}
\begin{equation}
\pi^{(2)}_{\rm DR} = \frac{8}{5}\,\Theta^{(2)}_2.
\label{anistropicstress}
\end{equation}
It is worth emphasizing that the anisotropic stress of free-streaming ultra-relativistic species is non-negligible.

Finally, we implement Eqs. (\ref{einsteinH}), (\ref{boltzmannhierarchy}), and (\ref{anistropicstress}) into the Einstein--Boltzmann solver \texttt{CLASS} \cite{class_1,class_2}\footnote{In practice, \texttt{CLASS} formulates the equations using a different variable, $F^{(m)}_\ell$, to describe the perturbations of the one-particle distribution function. The relation between $F^{(m)}_\ell$ and $\Theta^{(m)}_\ell$ is detailed in \cite{tram2013}.} to compute the evolution of cosmological perturbations.

\section{Impact on $C^{BB}_{\ell}$}\label{sec:impactoncls}

To evaluate the impact of self-interacting DR on the CMB we first study the efficiency of the interaction and then we compute the angular power spectrum of primordial $B$-mode polarization, using the \texttt{CLASS} implementation described in Sec. \ref{sec:modelimplementation} for different sets of parameters.

The efficiency of the interaction in an expanding universe is roughly determined by the ratio between the interaction rate $\Gamma$ and the Hubble parameter $H$. To be precise, the relevant ratio can be quantified by comparing the interaction term in Eqs. (\ref{boltzmannhierarchy}) with the Hubble parameter in physical units, i.e., $\alpha_2 \langle \Gamma \rangle / H$. Specifically, when $\alpha_2 \langle \Gamma \rangle/H \gg 1$, the interaction is efficient, whereas for $\alpha_2 \langle \Gamma \rangle/H \ll 1$, it is ineffective.

\begin{figure}
\includegraphics[width=\columnwidth]
{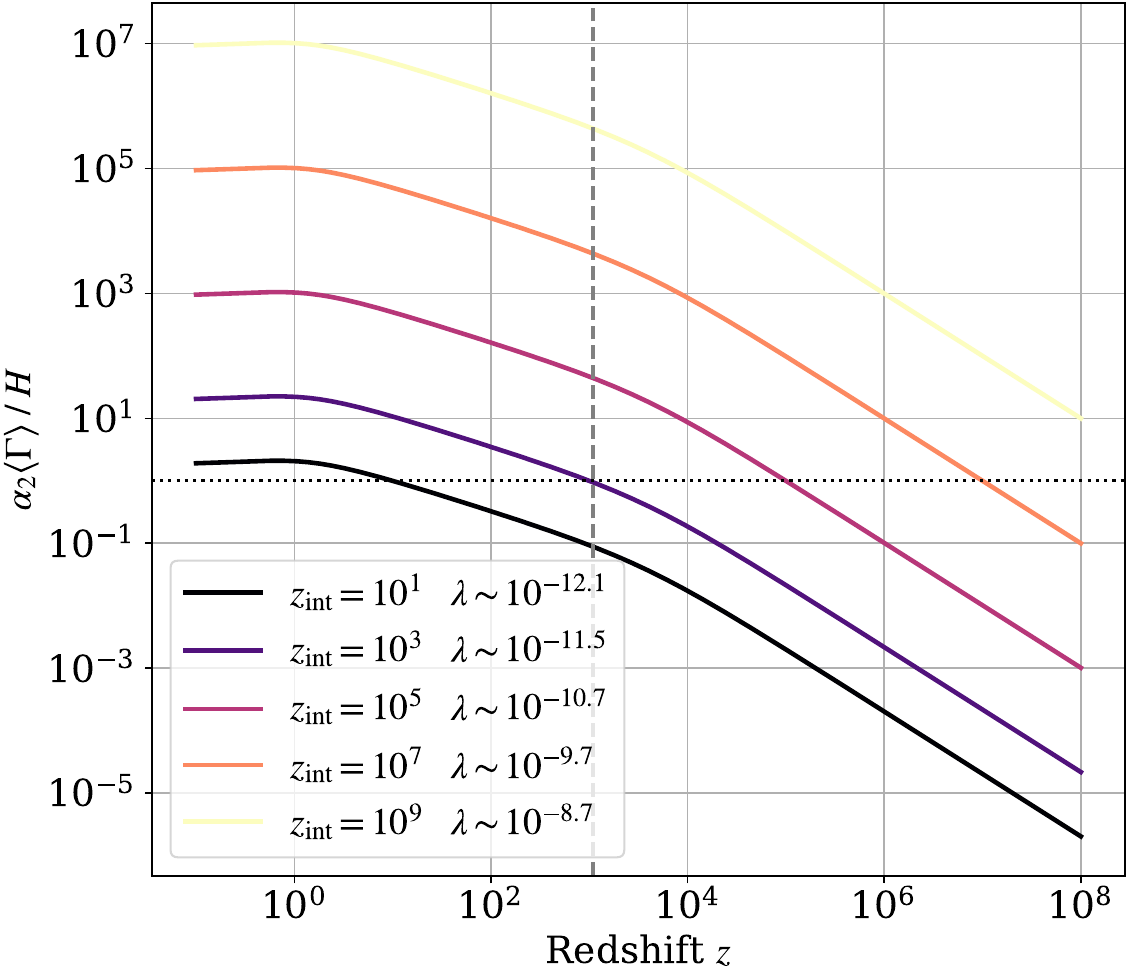}
\caption{\label{fig:gammaHint} Ratio of the DR self-interaction rate to the Hubble parameter as a function of redshift. Two distinct regimes are evident: the collisionless regime ($\alpha_2 \langle\Gamma\rangle/H < 1$), where DR free-streams and generates anisotropies, and the collision-dominated regime ($\alpha_2 \langle\Gamma\rangle/H > 1$), where self-interactions are efficient, driving the distribution function toward equilibrium and suppressing anisotropies. Since equilibrium is reached at late times, these interactions are referred to as recoupling. Solid curves show different self-interaction strengths, characterized by the recoupling redshift $z_{\rm int}$, or equivalently the self-coupling constant $\lambda$ (see Eqs. (\ref{gamma_lambda}) and (\ref{gammavsz})). The vertical gray dashed line indicates the recombination redshift $z_{\rm CMB}\sim1100$. For $z_{\rm int} > z_{\rm CMB}$, DR recouples before recombination, potentially leaving a noticeable imprint on the CMB, while for $z_{\rm int} < z_{\rm CMB}$, no significant effect is expected.}
\end{figure}

Following the approach of \cite{drint}, the self-interaction rate introduced in Eq. (\ref{gamma_lambda}) can be equivalently parametrized in terms of the redshift $z$ and a characteristic redshift $z_{\rm int}$, rather than the scale factor $a$ and the coupling constant $\lambda$. In this parametrization the interaction rate reads
\begin{equation}
\langle \Gamma \rangle = \frac{H_{\rm int}}{\alpha_2} \, \frac{1+z}{1+z_{\rm int}} \, ,
\label{gammavsz}
\end{equation}
where $z_{\rm int}$ is defined as the redshift at which $\alpha_2 \langle\Gamma\rangle = H$, and $H_{\rm int}\equiv H(z_{\rm int})$. The characteristic redshift $z_{\rm int}$ thus fixes the strength of the self-coupling $\lambda$ by equating Eqs. (\ref{gamma_lambda}) and (\ref{gammavsz}), and also determines the comoving relaxation time $\tau$ through Eq. (\ref{taugammalambda}).

Figure \ref{fig:gammaHint} displays the evolution of the ratio $\alpha_2 \langle \Gamma \rangle / H$, which quantifies the efficiency of self-interactions, as a function of redshift for different interaction strengths. By construction, the value of $z_{\rm int}$ determines the transition between the inefficient and efficient regimes. The scaling of the interaction rate, $\langle \Gamma \rangle \sim 1/a$, together with $H^{-1}\sim a^2$ during radiation domination and $H^{-1}\sim a^{3/2}$ during matter domination, implies that the ratio $\alpha_2 \langle \Gamma \rangle / H$ increases with time. Consequently, self-interactions can only become efficient at late epochs, a phenomenon referred to as recoupling, with $z_{\rm int}$ denoting the corresponding recoupling redshift.

\begin{figure}
\includegraphics[width=\columnwidth]{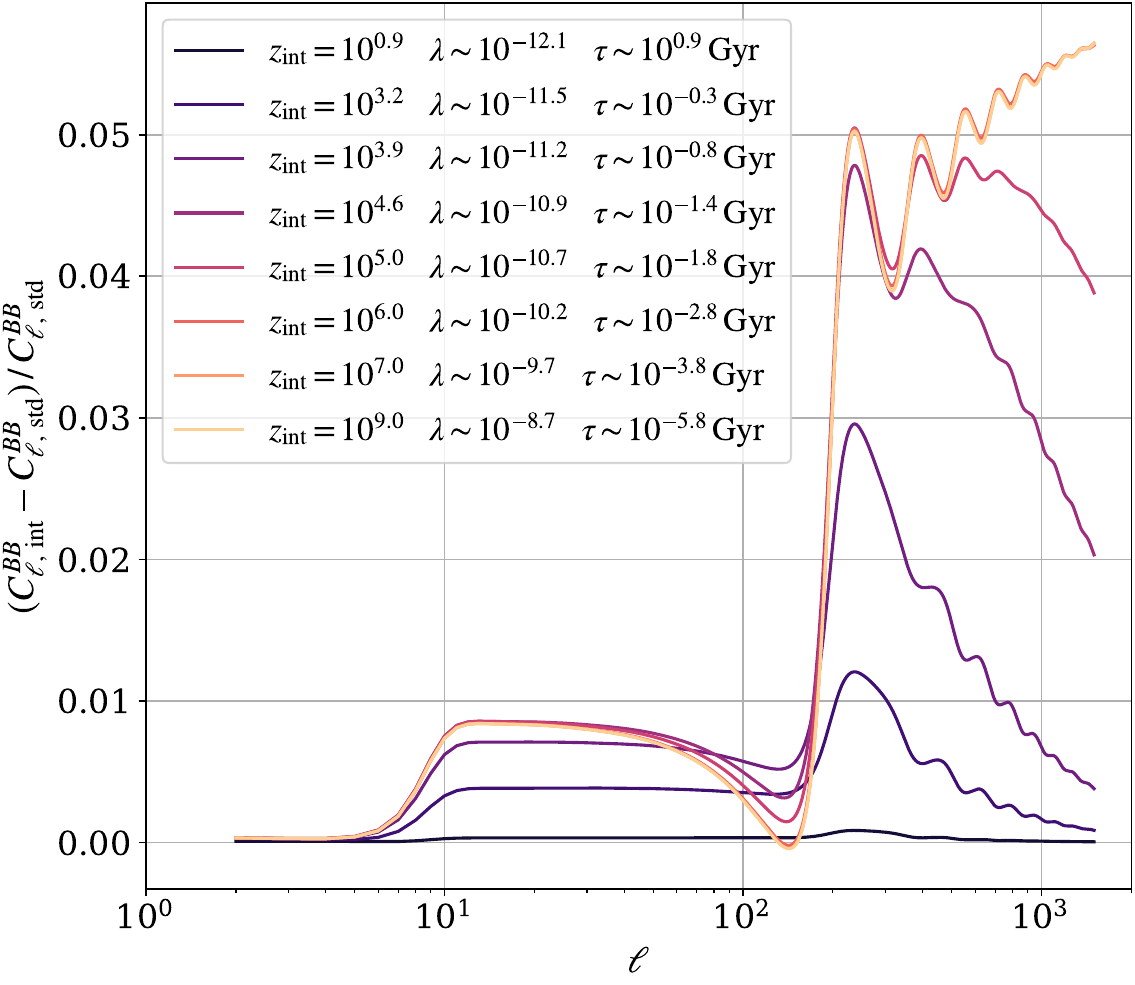}
\caption{\label{fig:relativediff}Relative difference in the primordial B-mode polarization angular power spectrum induced by self-interacting DR, compared to the standard free-streaming case, as a function of multipole moment $\ell$. Each curve corresponds to a different recoupling redshift $z_{\rm int}$, with the associated self-coupling $\lambda$ and interaction timescale $\tau$ indicated. Earlier interactions (i.e., larger $z_{\rm int}$, larger $\lambda$, or shorter $\tau$) result in stronger suppression of anisotropies due to efficient self-interactions. This erases the anisotropic stress responsible for GW damping, thereby enhancing the amplitude of primordial GW (see \cite{nahuel,baym,loverde2022}) and thus increasing the power in the B-mode spectrum. For sufficiently early interactions, $z_{\rm int}\gtrsim 10^7$, the effect saturates and no further enhancement is observed. Conversely, for late recoupling, $z_{\rm int}\lesssim z_{\rm CMB}\sim 10^3$, the impact is negligible since the self-interactions become effective only after recombination, when they can no longer influence the primary CMB anisotropies. For this plot, we assume $\Delta N_{\rm eff}=0.5$ as a representative example resulting in a roughly 5\% difference for larger $\lambda$. Decreasing the abundances to $\Delta N_{\rm eff}=0.1$ gives an effect of about 1\%.}
\end{figure}

At early times ($z > z_{\rm int}$), the ratio $\alpha_2 \langle \Gamma \rangle / H$ is smaller than unity, defining the collisionless regime. In this regime the self-interactions are inefficient and DR free-streams, thereby preserving its anisotropic stress (\ref{anistropicstress}). At later times ($z < z_{\rm int}$), the condition reverses and $\alpha_2 \langle \Gamma \rangle / H > 1$, which characterizes the collision--dominated regime. Here the self-interactions become efficient, driving DR into internal equilibrium suppressing its fluctuations.

Finally, varying the interaction parameters $\lambda$ or $\tau$ effectively shifts the recoupling redshift $z_{\rm int}$. Increasing $\lambda$ reduces the characteristic comoving interaction time $\tau$, which raises $z_{\rm int}$ and makes the interactions efficient at earlier epochs as shown in Fig. \ref{fig:gammaHint}.

The impact of self-interacting DR on the CMB is illustrated in Fig. \ref{fig:relativediff}, where we show the relative difference in the angular power spectrum of primordial $B$-mode polarization, computed using the modified version of \texttt{CLASS}, for various values of $\lambda$ (or, equivalently, $z_{\rm int}$) with respect to the standard case of non-interacting free-streaming DR ($\lambda=0$).

Although DR does not couple directly to CMB polarization, the two are linked gravitationally. In the tensor sector, which is the relevant one for primordial $B$ modes, the sequence of effects can be described as follows: a non-vanishing anisotropic stress [Eq. (\ref{anistropicstress})] damps the gravitational waves $h$ via Eq. (\ref{einsteinH}), and the evolution of these tensor perturbations in turn imprints on the CMB polarization, particularly the primordial $B$ modes. The amplitude of the anisotropic stress is governed by the Boltzmann hierarchy [Eq. (\ref{boltzmannhierarchy})] and depends sensitively on the efficiency of DR self-interactions, as discussed above.

The main effect of these interactions is an effective enhancement of the $B$-mode power spectrum amplitude across all multipoles in comparison to the standard free-streaming DR case, with the exception of $\ell \sim 100$. This enhancement grows with increasing coupling constant $\lambda$ (or, equivalently, higher $z_{\rm int}$). Earlier transitions (larger $\lambda$ or $z_{\rm int}$) correspond to stronger self-interactions that drive the one-particle distribution function of DR toward equilibrium more efficiently. In this regime, the anisotropic stress responsible for GW damping is erased, leading to a pronounced enhancement of power at high multipoles relative to the free-streaming case. This behavior highlights the fluid-like character of DR at late times. In contrast, if $z_{\rm int}<z_{\rm CMB}\sim 10^3$, the self-interactions become efficient only after recombination and, therefore, do not affect the primary CMB anisotropies.

As an illustrative example, Fig. \ref{fig:relativediff} shows the relative change in the $B$-mode spectrum for $\Delta N_{\rm eff}=0.5$. In the case of early recoupling (large $\lambda$), the variation reaches at most $\sim 5\%$. Reducing the abundance to $\Delta N_{\rm eff}=0.1$ suppresses the effect to the percent level ($\sim 1\%$). Since this is a relative difference, it is independent of the overall tensor amplitude $r$ and therefore applies to any $r \neq 0$. All other cosmological parameters are fixed to the Planck 2018 best-fit values \cite{planck18}.

\section{Impact on the tensor-to-scalar ratio}\label{sec:impactonr}

The observable imprint of self-interacting dark radiation (DR) on the cosmic microwave background (CMB) motivates a targeted assessment of whether neglecting DR interactions can bias cosmological-parameter inference --- in particular the tensor-to-scalar ratio $r$. To address this question we generate mock primordial (unlensed) CMB $B$-mode spectra using a modified version of \texttt{CLASS} that explicitly includes self-interacting DR. The injected (fiducial) cosmologies span different values of $\Delta N_{\rm eff}$ and of the self-coupling $\lambda$ (equivalently parametrized by $z_{\rm int}$ or $\tau$), so that the simulated data capture both the amplitude and the interaction-dependent perturbation phenomenology of DR.

These mock spectra include realistic uncertainties from instrumental noise, finite beam resolution and sampling (cosmic) variance. 
The covariance matrix of the binned $B$-mode angular power spectrum is taken in the standard diagonal form \cite{knox95},
\bea
{\rm Cov}_{\ell\ell'} \;=\; \left(\frac{1}{f_{\rm sky}\,\Delta\ell}\right)\frac{2}{2\ell+1}\Big[C_\ell + w^{-1} B_\ell^{-2}\Big]^2 \,\delta_{\ell\ell'},\label{eq:cov}
\eea
where the Gaussian beam transfer function is
\bea
B_\ell=\exp\!\Big[-\frac{\ell^2\theta_{\rm FWHM}^2}{16\ln 2}\Big],
\eea
with $\theta_{\rm FWHM}$ is the beam full-width at half-maximum, $w^{-1/2}$ is the total array Noise Equilavent Temperature level ($\mu$K-arcmin) of the maps (related to the observation time and the sensitivity and number of the detectors) and $f_{\rm sky}$ is the observed sky fraction. The bin width $\Delta\ell$ reflects our chosen multipole binning scheme (see below). The covariance is evaluated per bin and is diagonal in our setup, as indicated by the Kronecker delta $\delta_{\ell\ell'}$.

We adopt a binning scheme that ensures an equal number of modes per bin. Concretely, we set $N_{\rm bin}=100$ bins spanning $\ell_{\rm min}=10$ to $\ell_{\rm max}=1000$, such that each bin contains the same number of harmonic modes. To assess the dependence of our conclusions on the representative experimental sensitivity of next-generation CMB experiments, we consider two instrumental configurations based on the expected performance of the PICO mission \cite{2018SPIE10698E..46Y,Hanany2019,Aurlien_2023}. In both cases, we assume a sky coverage of $f_{\rm sky}=0.75$, while varying the beam width $\theta_{\rm FWHM}$ and the noise level to reflect 5- and 10-year observation periods, respectively. Specifically, we define \textbf{Noise 1:} as $\theta_{\rm FWHM} \simeq 1'$, $w^{-1/2} = 0.43~\mu{\rm K}\text{-arcmin}$ (10-year obs.), and \textbf{Noise 2:} as $\theta_{\rm FWHM} \simeq 2'$, $w^{-1/2} = 0.61~\mu{\rm K}\text{-arcmin}$ (5-year obs.). Together with the multipole binning scheme described above, these specifications fully determine the covariance matrix~\eqref{eq:cov}.

For each choice of fiducial cosmology based on the cosmological parameters fixed to the Planck 2018 best-fit values \cite{planck18} and specific $\Delta N_{\rm eff}$, interaction strength $\lambda$ and $r$, we simulate realizations of the primordial (unlensed) $C_\ell^{BB}$ including the instrumental noise described above. We then employ Markov Chain Monte Carlo (MCMC) analyses to recover cosmological parameters from each simulated dataset. To isolate the systematic effect of ignoring interacting DR on the inference of the tensor amplitude, we fit only the single parameter $r$ in all MCMC runs while holding the remaining cosmological parameters fixed at their fiducial values. This choice permits a direct comparison of statistical uncertainty (instrumental + cosmic variance) with the systematic shift induced by model misspecification.

To quantify the impact of self-interacting DR on parameter estimation    in a controlled way, we consider three combinations of input (true) and fitting models (see Table \ref{tab:typeofsim}):  
(i) \textbf{STD--STD}, where both the input and the fitting model assume free-streaming (standard) DR;  
(ii) \textbf{INT--INT}, where both input and fitting models include interacting DR; and  
(iii) \textbf{INT--STD}, where the input model contains interacting DR but the fitting model incorrectly assumes free streaming.  
(Hereafter we abbreviate ``interacting'' by INT and ``standard/free-streaming'' by STD.) For each case, we draw $N=500$ independent noise realizations, recover the best-fit $r$ for each realization via MCMC, and thereby construct the empirical posterior (sampling distribution) of the recovered $r$ values.

\begin{table}[b]
\caption{\label{tab:typeofsim}
Summary of the three analyses performed, combining different input and fitting models. The first two serve as consistency checks, where the same model is used for both input and fitting, thus no bias is expected. The third scenario is the main case of interest: the simulated universe includes self-interacting DR, but the fitting assumes standard, non-interacting DR to evaluate potential biases in estimating $r$.}
\begin{ruledtabular}
\begin{tabular}{ccc}
 Id& Input DR model& Fitting DR model\\
\hline
 STD-STD & Standard (no interac.) & Standard (no interac.)\\
 INT-INT & Self-interacting& Self-interacting\\
 INT-STD & Self-interacting & Standard (no interac.)\\
\end{tabular}
\end{ruledtabular}
\end{table}

This ensemble approach serves two closely related purposes. First, by aggregating results across many noise realizations we characterize the distribution (mean, variance, higher moments) of the recovered $r$ under each model pairing and instrumental configuration. Second, by comparing the distributions between matched and mismatched model pairs (INT--INT vs INT--STD, for example) we directly quantify the systematic bias in $r$ introduced when DR self-interactions are omitted from the fitting model, relative to the statistical uncertainty set by instrument noise and cosmic variance.

\begin{figure}
\includegraphics[width=\columnwidth]{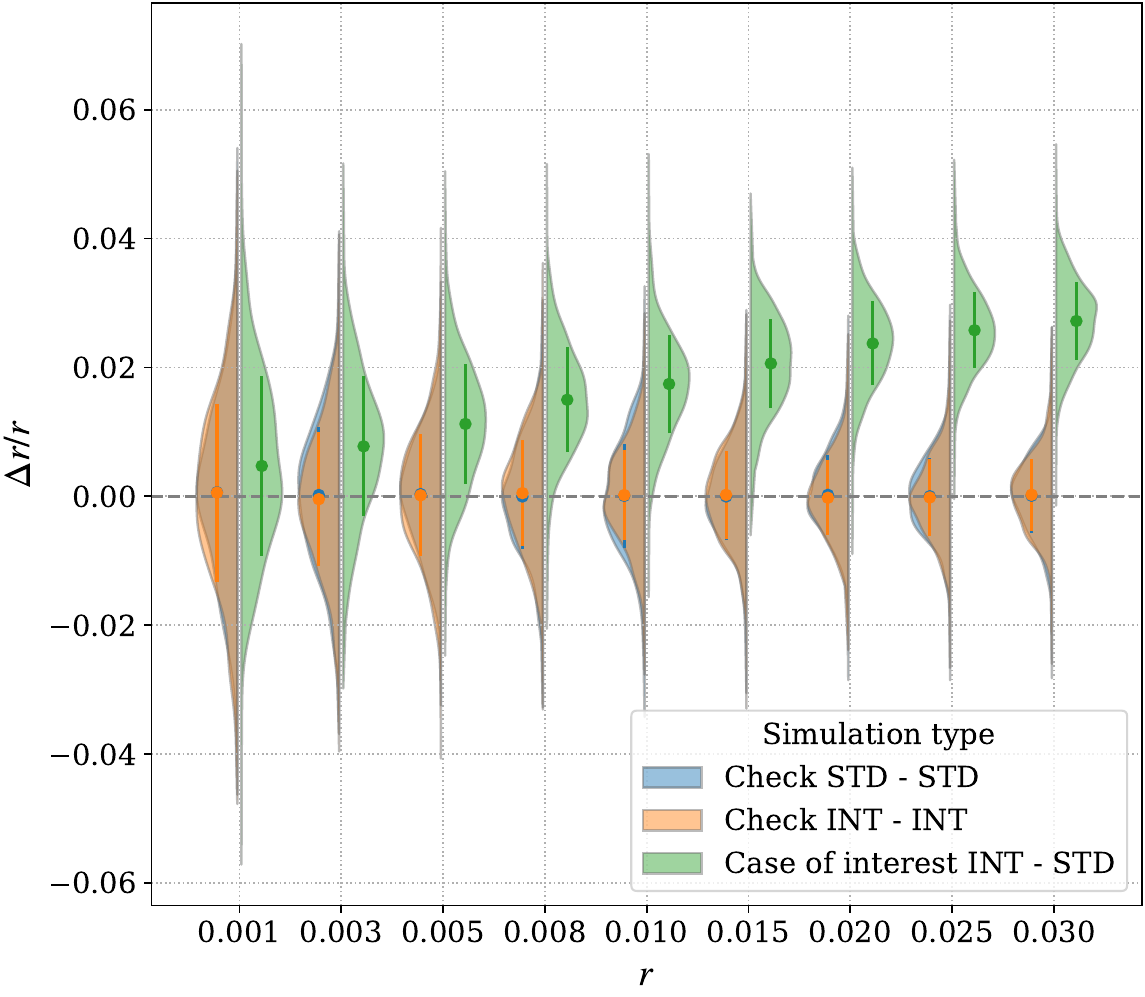}
\caption{\label{fig:violin}
Posterior distributions of the reconstructed relative bias of the tensor-to-scalar ratio, $\Delta r/r = (r_{\rm best-fit} - r_{\rm fid})/r_{\rm fid}$, obtained over 500 noise realizations for different fiducial values of $r$. Three types of simulated analysis described in Table \ref{tab:typeofsim} are shown: validation of the standard pipeline (Check STD-STD) in blue, validation of the interacting DR pipeline (Check INT-INT) in orange, and the main case of interest where self-interacting DR is analyzed assuming the standard model (INT-STD) in green. Colored dots and error bars represent the mean and the standard deviation, respectively. While the check cases exhibit negligible bias across the $r$ range, the INT-STD case shows a systematic overestimation of $r$ illustrating the bias introduced when self-interacting DR is incorrectly modeled as standard free-streaming.} 
\end{figure}

Figure \ref{fig:violin} shows the posterior distribution of the relative bias,
\bea
\frac{\Delta r}{r} \equiv \frac{r_{\rm best\!-\!fit}-r_{\rm fid}}{r_{\rm fid}},
\eea
as a function of the fiducial value $r_{\rm fid}$. We restrict to $r_{\rm fid}<0.03$, consistent with current observational bounds \cite{bicepkeck,Tristram_2022} (see also \cite{neff-act25}). The results are obtained from $500$ independent noise realizations, using a representative choice of DR parameters with $\Delta N_{\rm eff}=0.5$ and $z_{\rm int}=10^7$ (corresponding to $\lambda\simeq10^{-10}$). The STD--STD and INT--INT cases are consistency checks that validate the numerical pipeline, in both the fiducial input and fitting models match and the recovered distribution of $\Delta r$ do not show any appreciable bias within statistical expectations. In contrast, the INT--STD case, for which the input fiducial cosmology contains interacting DR but the fitting model assumes standard free streaming, indicates a clear systematic overestimation of $r$, i.e. the inferred tensor amplitude is consistently biased high.

\begin{figure}
\includegraphics[width=\columnwidth]{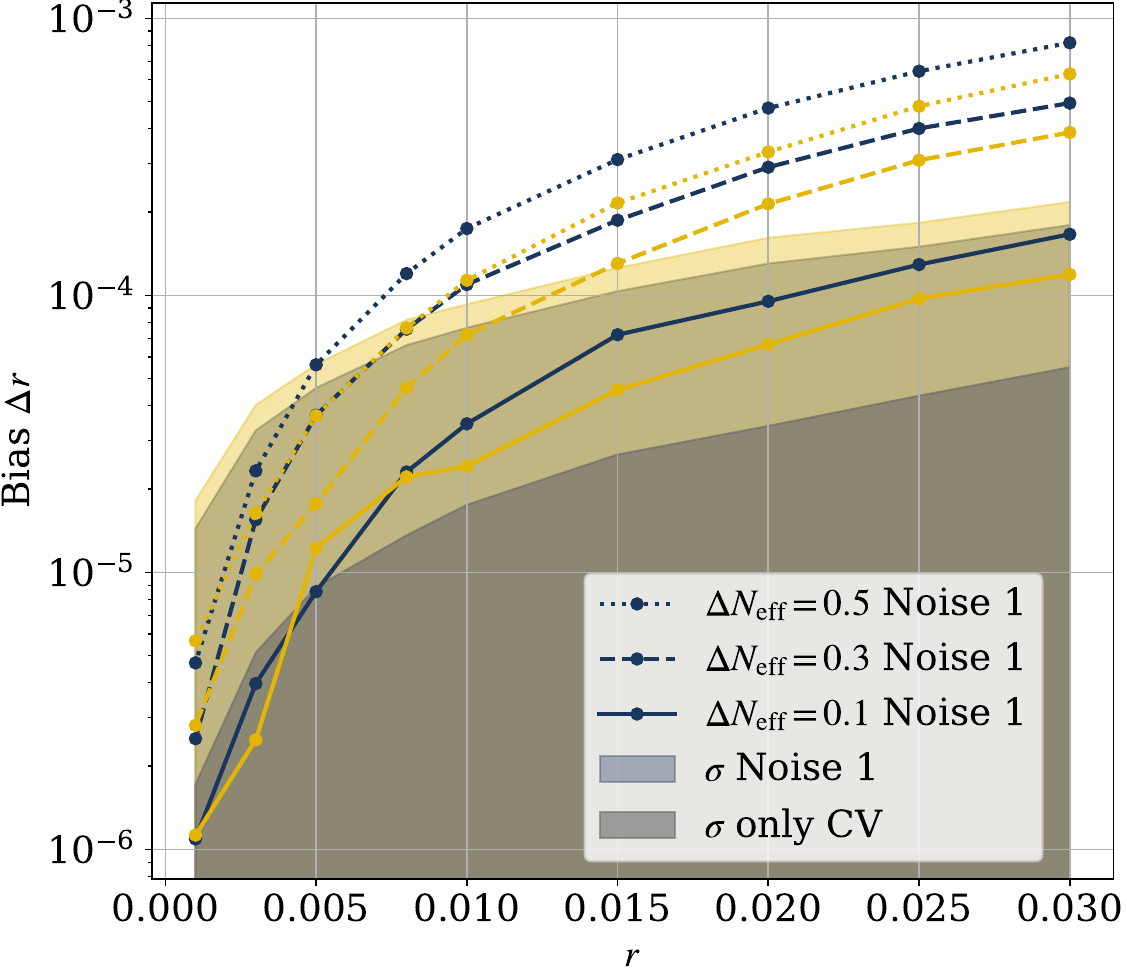}
\caption{\label{fig:biasvsnoise}Bias in the reconstructed tensor-to-scalar ratio $\Delta r=r_{\rm best-fit}-r_{\rm fid}$, as a function of the fiducial value of $r$, for different values of $\Delta N_{\rm eff}$, $z_{\rm int} \simeq 10^7$ ($\lambda\simeq10^{-10}$), and two experimental noise configurations: Noise 1 (dark blue) and Noise 2 (yellow) representative of next-generation CMB experiments. Solid, dashed, and dotted lines correspond to $\Delta N_{\rm eff} = 0.1$, $0.3$, and $0.5$, respectively. The shaded regions indicate the standard deviation related to the posterior distribution in Fig. \ref{fig:violin} for Noise 1 (dark blue), Noise 2 (yellow), and a cosmic variance–limited scenario (only CV in dark gray). The results show that the bias can be comparable to or even exceed the statistical uncertainty, underscoring the need to accurately model DR interactions in future precision measurements of primordial B-modes. In this analysis, we isolate the impact of self-interacting DR on the estimation of $r$ by considering only this specific systematic effect, excluding others.}
\end{figure}

\begin{figure}
\includegraphics[width=\columnwidth]{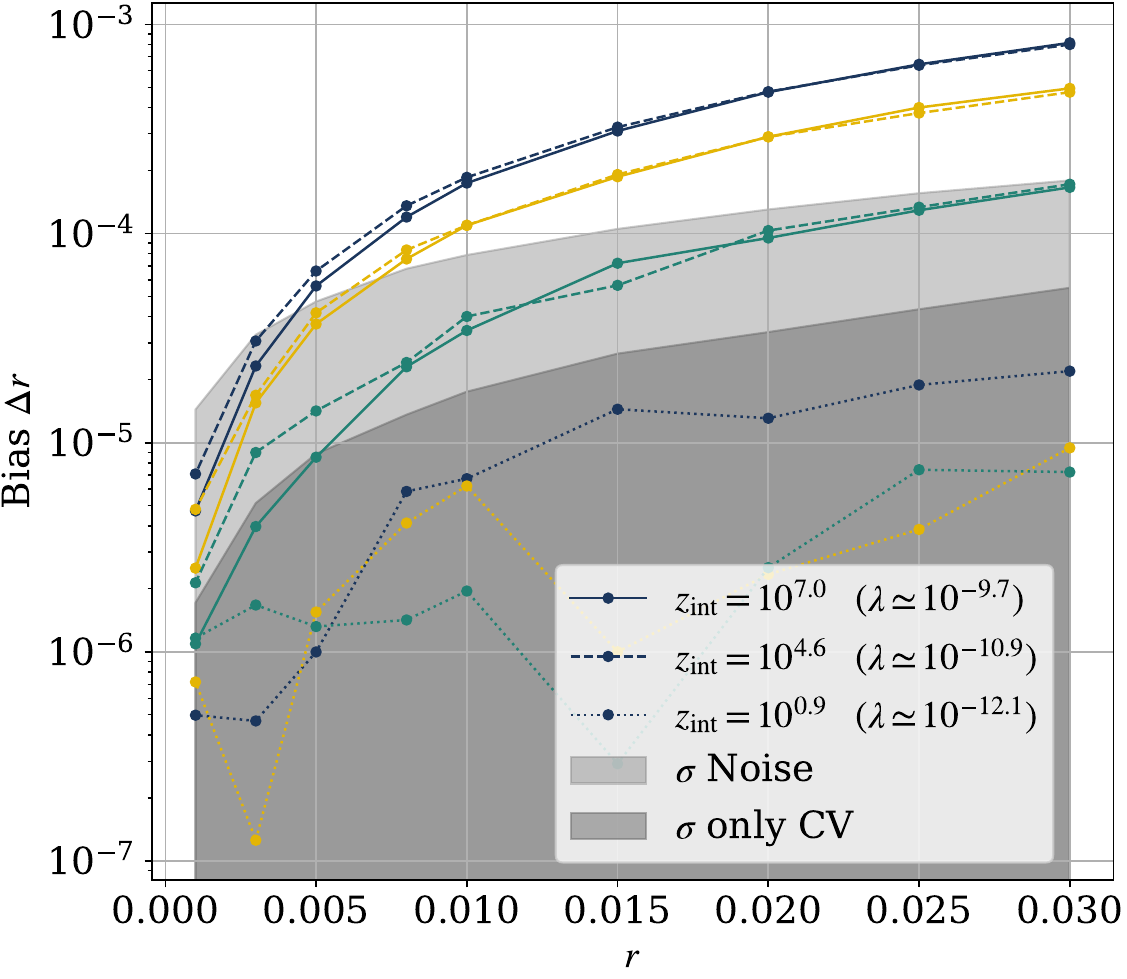}
\caption{\label{fig:biasvslambda}
Bias in the recovered tensor-to-scalar ratio, $\Delta r$, as a function of the fiducial $r$. Different line styles indicate DR self-interaction strengths, parametrized by the recoupling redshift $z_{\rm int}$ (or equivalently, the self-coupling $\lambda$). Colors correspond to DR abundance: $\Delta N_{\rm eff}=0.5$ (dark blue), $\Delta N_{\rm eff}=0.3$ (yellow), and $\Delta N_{\rm eff}=0.1$ (green). Three representative interaction regimes are shown: early ($z_{\rm int} \simeq 10^7$), intermediate ($z_{\rm int} \simeq 10^{4.6}$), and late ($z_{\rm int} \simeq 10^{0.9}$) recoupling. Shaded regions represent the $1\sigma$ statistical uncertainty on $r$ for the experimental configuration Noise 1 (light gray) and the cosmic variance–limited case (only CV in dark gray). As expected from the behavior shown in Fig. \ref{fig:relativediff}, the bias saturates for early interactions ($z_{\rm int}\gg z_{\rm CMB}\sim 10^3$, i.e., large $\lambda$) and it is comparable to the $1\sigma$ level, while it becomes negligible for late interactions ($z_{\rm int}\lesssim z_{\rm CMB}$, i.e. small $\lambda$), where DR interactions become efficient too late to affect the primordial B-mode signal.}
\end{figure}

To quantitatively assess this systematic bias effect on the tensor-to-scalar ratio $r$ observed in the INT--STD case where the input self-interacting DR is incorrectly modeled as standard free-streaming radiation, we compare the magnitude of the bias with the maximum standard deviation $\sigma$ derived from the posterior distributions in Figure \ref{fig:violin} for each $r$. Figure \ref{fig:biasvsnoise} presents this comparison for different values of $\Delta N_{\rm eff}$ and a fixed, strong self-interaction strength, $\lambda \simeq 10^{-10}$ (or $z_{\rm int} \simeq 10^7$). The bias and the $\sigma$ uncertainty level are shown for the two experimental noise configurations mentioned above: Noise 1 and Noise 2. As expected, the effect vanishes in the limit $\Delta N_{\rm eff}\to 0$, but for values as small as $\Delta N_{\rm eff}\sim 0.1$ the induced systematic already surpasses the cosmic variance level and becomes comparable to $\sigma$ particularly in the range $r \gtrsim 5 \times 10^{-3}$. This highlights the need to account for such effects in future high--precision measurements of primordial $B$-modes and of the tensor--to--scalar ratio $r$.

Finally, we investigate the dependence of this bias on the strength of the DR self-interaction, parameterized by the self-coupling constant $\lambda$ or, equivalently, the recoupling redshift $z_{\rm int}$. Figure \ref{fig:biasvslambda} shows the results for the `Noise 1' experimental configuration. For fixed $\Delta N_{\rm eff}$, the key finding is that the bias is significant only if the DR recoupling occurs before recombination, i.e. $z_{\rm int} > z_{\rm CMB} \sim 10^3$ (see Figure \ref{fig:gammaHint}). In this regime, the bias increases with the strength of the interaction and saturates for early interactions $z_{\rm int} \gtrsim 10^7$ (or $\lambda\gtrsim 10^{-10}$). Conversely, for interactions that become efficient only after recombination, i.e. $z_{\rm int} \lesssim 10^3$ or $\lambda< 10^{-11}$, the impact on the primary CMB anisotropies is negligible, and no appreciable bias on $r$ is introduced. This whole behavior is consistent with the trends already identified in Figure \ref{fig:relativediff}, where we quantify the impact of these interactions on the primordial $C_\ell^{BB}$.

\section{Discussion and conclusions}\label{sec:conclusions}

We have investigated the cosmological consequences of self-interacting DR with particular emphasis on its imprint on primordial $B$-mode polarization and on the estimation of the tensor-to-scalar ratio $r$. Using a minimal benchmark in which DR is modeled as an effectively massless scalar with a quartic self-interaction $\lambda\phi^4/4!$, we implemented the corresponding modifications to the Einstein--Boltzmann system in a modified version of \texttt{CLASS} and explored observationally motivated regions of parameter space (e.g. $\Delta N_{\rm eff}\lesssim 0.5$ and $\lambda\sim10^{-12}\!-\!10^{-9}$, or equivalently recoupling redshifts $z_{\rm int}$ spanning early to late epochs).

Our main technical result is that DR self-interactions change the tensor-sector transfer functions in a scale-dependent way. Physically, when self-interactions become efficient (``recoupling'' at $z\lesssim z_{\rm int}$) the DR fluid is driven toward local equilibrium and its anisotropic stress is suppressed. Because anisotropic stress produced by free-streaming relativistic species is the source of gravitational-wave damping, efficient self-interactions reduce that damping and therefore enhance the primordial $B$-mode power relative to the free-streaming case. The effect is most pronounced at angular scales corresponding to modes that re-enter the horizon around or after the recoupling epoch and grows with the DR abundance $\Delta N_{\rm eff}$ and with the strength of the interactions, larger $\lambda$ (or equivalently earlier recoupling, larger $z_{\rm int}$).

We quantified this behaviour in two complementary ways. First, we computed the relative modification of the primordial (unlensed) $B$-mode spectrum across multipoles for representative $(\Delta N_{\rm eff},\lambda)$ choices and identified the multipole dependence of the effect in Sec. \ref{sec:impactoncls}. Second, to assess observational impact we produced realistic mock $B$-mode datasets (including instrumental noise based on the features of the forthcoming CMB polarization experiments and cosmic variance) and performed MCMC parameter recovery with fitting models that either include or neglect DR self-interactions (see Sec. \ref{sec:impactonr}). By construction, our analysis isolates the systematic error on $r$ produced when DR interactions are modeled incorrectly.

Our results show that the systematic bias in the recovered tensor amplitude can be non-negligible. For parameter choices near current upper bounds (e.g., $\Delta N_{\rm eff}\sim 0.1$--$0.5$ and early recoupling), the inferred $r$ may be overestimated by an amount comparable to, or even exceeding, the statistical sensitivities forecasted for forthcoming CMB experiments. The projected sensitivities at the level $\sigma(r)\sim10^{-3}$ for experiments such as \textit{The Simons Observatory} and \textit{LiteBIRD}, and the still more ambitious targets ($\sigma(r)\lesssim \mathcal{O}(1)
\times10^{-4}$) envisaged for concepts like CMB-S4/PICO \cite{inflationtheoryobservations,Abazajian2022,Ade2019,so_2025,Allys2022,2018SPIE10698E..46Y,cmbs4extended,cmbs4finalreport,Abazajian2022,Hanany2019,Aurlien_2023,giardiello_requirements_2025}, highlight the observational relevance of this bias in the near future. Figures \ref{fig:biasvsnoise} and \ref{fig:biasvslambda} illustrate this result by showing that, for moderate tensor amplitudes $r \gtrsim 5 \times 10^{-3}$, the bias $\Delta r$ can reach the level of representative noise across different values of $\Delta N_{\rm eff}$, $\lambda$ (or $z_{\rm int}$), and experimental configurations.

From the standpoint of data analysis, our results motivate the inclusion of parametrized models of DR self-interactions in likelihood pipelines used to infer $r$; to jointly constrain $\Delta N_{\rm eff}$ and interaction parameters rather than assuming a free-streaming DR model; and quantify the sensitivity of $r$ constraints to unmodeled interactions as part of systematic-error budgets for next-generation surveys. In particular, the bias estimated in this work, $\Delta r \gtrsim 10^{-4}$, exceeds the \textit{LiteBIRD} budget allocation per systematic effect ($\Delta r_{\rm sysLB} \simeq 6.5\times10^{-6}$)~\cite{giardiello_requirements_2025}.

Several caveats are important. Our conclusions are drawn within a specific DR model; other interaction forms, additional scattering channels, or different particle content could modify the detailed multipole dependence and amplitude of the effect. Moreover, degeneracies with other cosmological parameters (e.g. foreground residuals, delensing efficiency, or uncertainties in $\Delta N_{\rm eff}$) may weaken or amplify the observable bias, and a full end-to-end analysis including foregrounds and instrumental systematics would be required to assess the accurate impact on a given experiment.

As future work, it will be important to explore a wider class of interaction operators and particle models (fermionic DR, scalar and vector mediators), study degeneracies with foreground modeling and delensing residuals in realistic pipeline-level forecasts, and assess complementarity with non-CMB probes, such as Big Bang nucleosynthesis, large-scale structure, and laboratory/astrophysical searches, that can independently constrain $\Delta N_{\rm eff}$ and interaction strengths. Addressing these points will refine the criteria for when DR interactions must be modeled to avoid biased inferences about primordial GW.

In summary, self-interacting DR represents a plausible effect that can leave an observable imprint on primordial $B$ modes and, if neglected, induce a bias in the recovered tensor-to-scalar ratio at a level relevant for upcoming and future CMB experiments. Accounting for this effect will make detections of primordial GWs more robust and enhance the reliability of cosmological conclusions about the early-universe physics drawn from $B$-mode measurements.

\begin{acknowledgments}
NMG acknowledges financial support from CONICET Grant No. PIP2017/19:11220170100817 and from Universidad de Buenos Aires through Grant No. UBACYT 20020170100129BA. C.G.S. acknowledges funding from CONICET (PIP-2876), and Universidad Nacional de La Plata (G11-175), Argentina. 
\end{acknowledgments}


\bibliographystyle{apsrev4-2}  
\bibliography{references}       

\end{document}